# Enhanced nuclear fusion in the sub-keV energy regime

Micah. E. Karahadian[1], Matthew Colborne[2], Arun Persaud[2], Thomas Schenkel[2†], Jeremy N. Munday[1*]

[1]Department of Electrical and Computer Engineering, University of California, Davis, One Shields Avenue, Davis, CA 95616-5270, USA.
[2]Accelerator Technology and Applied Physics Division, Lawrence Berkeley National Laboratory, 1 Cyclotron Road, Berkeley, CA 94720, USA.
[†]email: t_schenkel@lbl.gov
[*]email: jnmunday@ucdavis.edu

## Abstract:

Nuclear fusion requires overcoming or traversing a repulsive Coulomb barrier of hundreds of kiloelectronvolts, rendering the probability of fusion at sub-keV energies vanishingly small. Yet in condensed matter, the electronic and structural environment of reacting nuclei can profoundly alter fusion rates. Here we demonstrate that deuterium–deuterium fusion within metallic foils exhibits a pronounced enhancement and reaction yield plateau below energies of 2.5 keV—contrary to the expected exponential suppression with decreasing energy. Using a dual-chamber platform that combines electrochemical deuterium loading with ion-beam bombardment, we show that fusion yields in palladium and titanium hydrides are enhanced by over $10^{18}$ compared to theoretical bare-nucleus fusion rates. These results demonstrate that access to low-energy fusion processes can be governed by materials degrees of freedom. This materials-driven fusion regime establishes a reproducible, tunable framework for studying and ultimately engineering nuclear reactions in solids. While the reaction rates reported here are low, these insights into materials-modulated fusion processes offer a potential foundation for understanding how condensed-matter environments could influence future fusion-energy concepts.

## Main Text

Nuclear fusion is the energy source of stars and a long-standing aspiration for terrestrial power generation. For light-nuclei fusion, Coulomb repulsion imposes a 100s of keV-scale barrier. Quantum mechanical tunneling allows for ions to fuse without the need to overcome this barrier; however, reaction rates fall exponentially with decreasing ion energy or temperature[1]. Progress toward reliable control of fusion plasmas for power production has accelerated in recent years with a series of promising confinement approaches[2–5]; though, many challenges remain[6–8]. In parallel, there is renewed interest in probing fusion in solids, where mechanisms exist to enhance the fusion rates[9-11].

Although the fusion rates accessible in condensed-matter systems are currently too small to be energetically useful, understanding how electronic structure, defect landscapes, and hydrogen transport modify tunneling probabilities is increasingly important for the broader fusion-energy community. Materials-driven modulation of nuclear reaction environments could help constrain models of screening, guide the design of enabling components for future fusion devices, and clarify whether solid-state environments provide regimes of fusion behavior not accessible in plasmas.

Ion beam experiments with solid targets provide a controlled means to study fusion at low reaction energies, where the electronic structure of the host material influences the reaction environment. Instead of colliding with bare nuclei, projectile ions interact with target nuclei embedded in materials, whose electron clouds can partially shield the nuclear charge. This phenomenon, referred to as electronic screening, lowers the effective Coulomb barrier, allowing nuclei to approach more closely and thereby enhance the tunneling probability compared with unscreened reactions. In metal hydrides, electron screening is parameterized by a screening potential, $U_e$, which is a correction to the reaction energy applied as an effective energy shift, $E_{eff} = E + U_e$, which reduces the kinetic energy needed by the ions to fuse[1,12]. Values of $U_e$ vary across materials from a few tens to several hundred eV[13–16].

Prior reports point to the potential for fusion rate modification at low energies. Chen *et al.* demonstrated that electrochemical loading of deuterium (D) into a Pd foil increases the fusion yield by 15-18% at center-of-mass (cm) energies down to $E_{cm}$ = 15 keV[10]. This experiment showed that meV-scale electrochemistry can affect MeV-scale fusion rates through an increase in the deuterium density within the Pd lattice but was unable to observe the effects of electron screening, which are proposed to occur at lower energies[12,17,18]. At low energies, the complex interplay between electron screening, deuterium density, and ion energy requires a different approach where sub-keV deuterium ions can be used to systematically interrogate fusion rates across dissimilar hydrides under varied deuterium loading conditions.

Here we report experiments that combine very low-energy (<1 keV) deuteron beams with electrochemical loading of deuterium into palladium and titanium foils and discover significant fusion rate enhancements. D-D fusion rates were quantified across $E_{cm}$ = 0.25 – 6.5 keV by detecting 3.02 MeV protons and 2.45 MeV neutrons. The combined action of deuterium ion implantation and electrochemical loading yields fusion rates nearly $10^9$ times higher than predicted for screened reactions and more than $10^{18}$ above standard theoretical cross-section expectations[19]. These results suggest a synergistic mechanism in which low-energy ions provide momentum and generate lattice defects, while electrochemical loading supplies mobile deuterium and additional vacancies[20,21]. Together, these effects sustain anomalously high fusion rates at sub-keV energies, opening new pathways to probe and optimize fusion at the lowest reaction energies measured to-date.

## Experimental set-up

Experiments were performed with palladium and titanium foils (250 μm thick) serving as membranes between an electrochemical cell and a high-vacuum ion-beam chamber (Fig. 1a). Palladium, with its high deuterium mobility and near-unity terminal solubility (D/Pd ≤ 1)[22], provides a benchmark system where strong electron screening has long been reported[23,24]. Titanium, by contrast, despite a higher solubility limit (D/Ti ≈ 2), exhibits lower diffusivity and is widely regarded as weakly screening[14,16,25].

Electrochemical loading of deuterium was carried out in a heavy-water electrolyte (6.0M $D_2SO_4$ in $D_2O$) under galvanostatic control, driving deuterons from solution into the metal lattice. To facilitate efficient chemisorption in Ti, the electrolyte-facing surface was coated with a thin Pd layer. A pressure transducer and residual-gas analyzer tracked permeating hydrogen in real time, where pressure increases linearly with applied current density (Extended Data Fig. 1). On the vacuum side, the same foil was simultaneously bombarded with a monoenergetic deuteron beam produced by a microwave-driven ion gun, tunable between $E_{cm}$ = 0.25 keV and 6.5 keV. Comparative chemical, structural, and morphological analyses of pristine and irradiated targets were carried out and are consistent with the expected degradation associated with ion beam bombardment (Extended Data Fig. 2). Nuclear fusion products were monitored in both channels of the D-D reaction: 2.45 MeV neutrons with a liquid scintillator coupled to pulse-shape-discriminating electronics (Fig. 1b) and 3.02 MeV protons with a passivated implanted planar silicon diode (Fig. 1c). Detector calibrations, beam flux normalization, and detailed descriptions of the electrochemical and ion-beam systems are provided in Methods. This configuration allows direct comparison between beam-only and beam-plus-electrochemical loading conditions, enabling systematic measurements of screening potentials across two canonical hydrides (PdD and TiD) in an unprobed energy regime where electron screening is expected to dominate fusion dynamics.

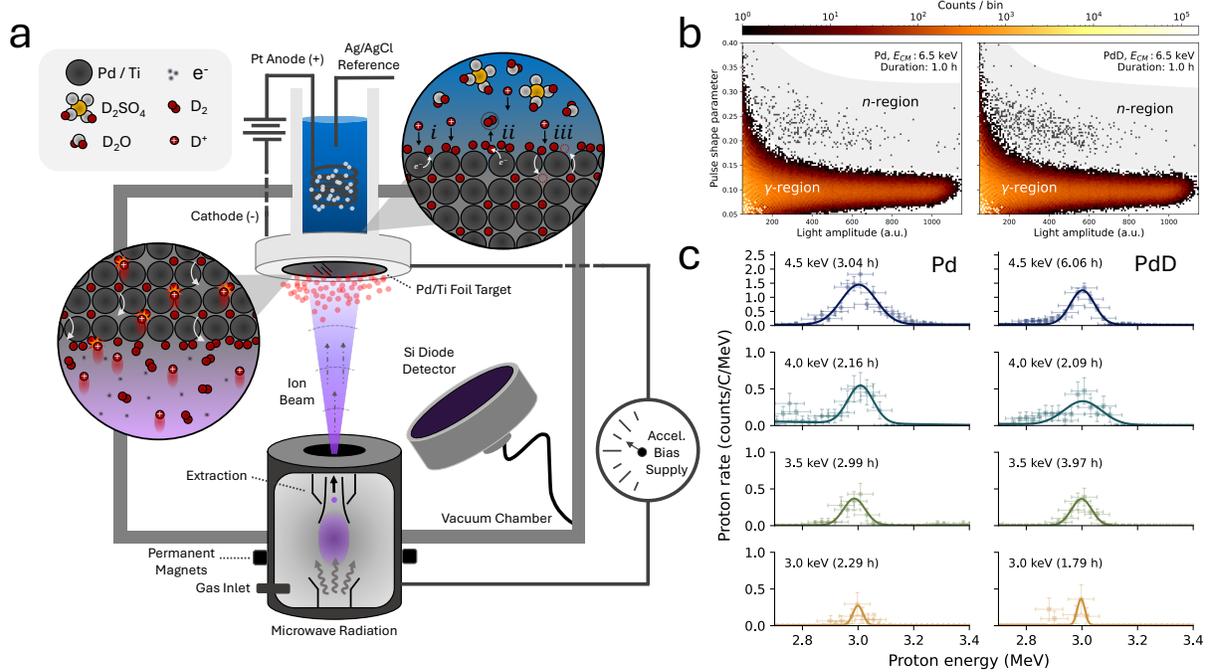

**Figure 1 | Experimental platform for probing sub-keV fusion in metal hydrides. (a)** Schematic showing the experiment configuration. A Pd or Ti foil separate an electrochemical cell (top) from a high-vacuum ion-beam chamber (bottom), enabling simultaneous electrochemical loading and low-energy beam implantation ($E_{cm}$ = 0.25 – 6.5 keV). On the electrolyte side, interfacial deuterium coverage and uptake are governed by the canonical (*i*) Volmer, (*ii*) Tafel/Heyrovský, and (*iii*) subsurface exchange (see Methods). Charged particles (notably 3.02 MeV protons) are recorded with a silicon diode; neutrons are detected with an organic liquid scintillator coupled to a photomultiplier (not shown). **(b)** Pulse-shape-discrimination (PSD) maps of scintillator light amplitude and pulse shape parameter at $E_{cm}$ = 6.5 keV (1 h runs) for Pd (left) and PdD (right), where PdD denotes the use of electrochemistry to insert deuterium into the metal lattice in addition to ion bombardment. The lower band corresponds to γ-ray events and the upper band to neutron events; the shaded region indicates the neutron selection used for yield extraction. **(c)** Proton energy spectra from the silicon diode for $E_{cm}$ = 4.5 – 3.0 keV, shown for Pd (beam-only, left column) and PdD (beam and electrochemical loading, right column). Rates are reported as counts/C/MeV (scaled by the measured beam current), and the ∼3 MeV peak is evident at each energy. A constrained fit (solid line) is shown as a guide to the eye.

# Results and Analysis

The rate of detected fusion products (3.02 MeV protons and 2.45 MeV neutrons) strongly depends upon the incident deuterium ion energy and separates into two energy regimes with qualitatively different behaviors: a screening region ($E_{cm}$ = 2.5 – 6.5 keV) and a lower-energy plateau region ($E_{cm}$ = 0.25 – 2.5 keV). The plateau indicates the emergence of additional effects within this reaction energy regime. Figure 2 shows the relative fusion yields for metal targets upon ion bombardment alone (Pd and Ti) and with electrochemical loading of deuterium (PdD and TiD) across both regimes. In these experiments, the detected proton signals dropped to background levels during electrochemical loading in the absence of ion beams. Proton and neutron counts are scaled per Coulomb using Faraday-cup measured currents and are normalized to each material's highest-energy reference point ($E_{cm}$ = 6.5 keV) to achieve the relative yield. This procedure (details in Methods, Extended Data Fig. 3) enables direct comparisons between metals and their deuterides and consistent extraction of the effective screening potential $U_e$ for each material.

Fitting the relative yields with screened cross-section models, we find that the effective screening potential in Pd increases from $U_e = 0.8 \pm 0.2$ keV in the beam-loaded foil to $1.7 \pm 0.2$ keV with beam plus electrochemical loading. For Ti, the corresponding increase is from $U_e = 0.3 \pm 0.2$ keV to $0.9 \pm 0.2$ keV. Thus, while electrochemical loading introduces a similar increment of approximately 0.6 to 0.9 keV to the apparent screening potential, the absolute magnitude remains much larger in the Pd system.

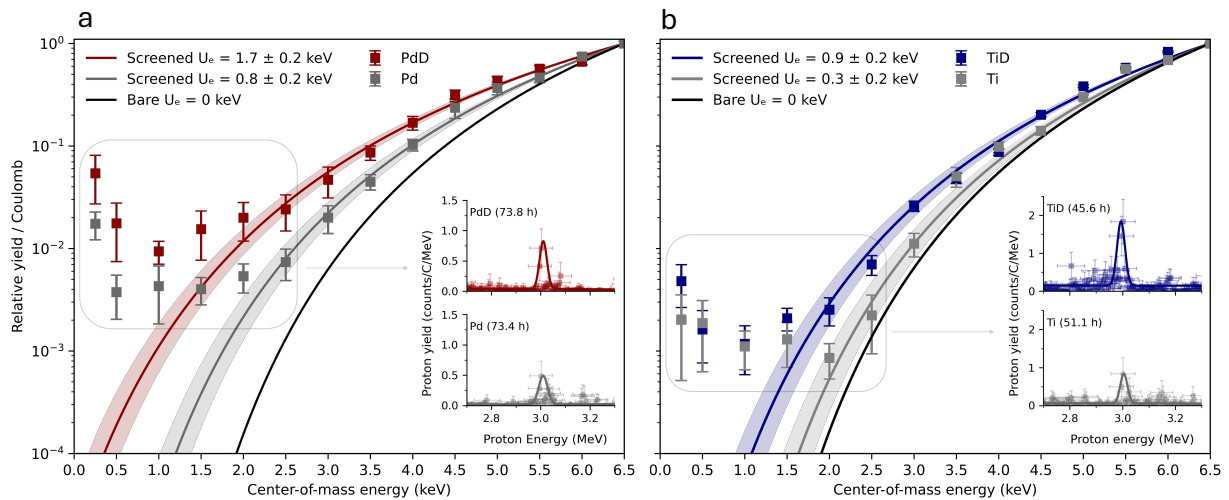

**Figure 2 | Relative thick-target yields for Pd, PdD, Ti, and TiD. (a)** Relative yields for beam-only Pd (gray) and beam with electrochemically loaded PdD (red). Curves are thick-target yield calculations using a screened cross-section with constant screening potential $U_e$ (red and gray) and the unscreened, "bare", case (black), $U_e = 0$. All datasets are normalized to the highest-energy point (6.5 keV, center-of-mass). Fits to the data from 6.5 keV to 2.5 keV give screening potentials of $U_e = 0.80 \pm 0.2$ keV (Pd) and $U_e = 1.70 \pm 0.20$ keV (PdD). Below 2.5 keV, a yield plateau and slight increase is observed for the lowest ion beam energies. **(b)** Relative yields for Ti (gray) and TiD (blue). Electrochemical loading increases the screening potential from $U_e = 0.3 \pm 0.2$ keV (Ti) to $U_e = 0.9 \pm 0.2$ keV (TiD) for the fitted data at energies >2.0 keV, while the relative yields for Ti and TiD remain comparable within the low-energy plateau (< 2.5 keV). Insets: cumulative 3 MeV proton spectra integrated over the shaded window (0.25–2.5 keV); total acquisition times are noted. Poisson counting uncertainties are propagated through the normalization and solid lines are included as a guide to the eye.

We compare the fusion rate in the metal foils to the expected fusion rate of bare, unscreened nuclei[19] by defining a fusion enhancement factor, which is the ratio of the measured screened yield to the bare (unscreened) yield at the same center-of-mass energy: $\mathcal{E}(E) = Y_{scr}(E)/Y_{bare}(E)$. Figure 3 shows $\mathcal{E}(E)$ for (a) Pd and PdD and (b) Ti and TiD. The dotted curves show the behavior we would expect from a parameter-free model of screened yield (with screening potentials taken from fits in the energy region of 2.5-6.5 keV); however, across both material systems this model fails below $E_{cm}$ = 2.5 keV. In contrast, a bi-exponential model (solid curves) provides a markedly improved fit (also capturing the screening effect at energies above 2.5 keV, not shown). At the lowest reaction energies in our experiments, the fusion enhancement factor approaches or exceeds $10^{18}$ for all materials considered (Pd, PdD, Ti, and TiD). Figure 3c shows the hydride-to-metal relative yield ratio, $Y_{hydride}/Y_{metal}$, for both material systems showing that electrochemical loading increases the yield in both metal hosts and trends toward unity at higher energies, as expected. Across the measured ranges, the relative yield ratio for both Pd and Ti are broadly similar and overlap within uncertainties. Proton- and neutron-based ratios show similar enhancements, indicating that the loading effect is consistent across diagnostics.

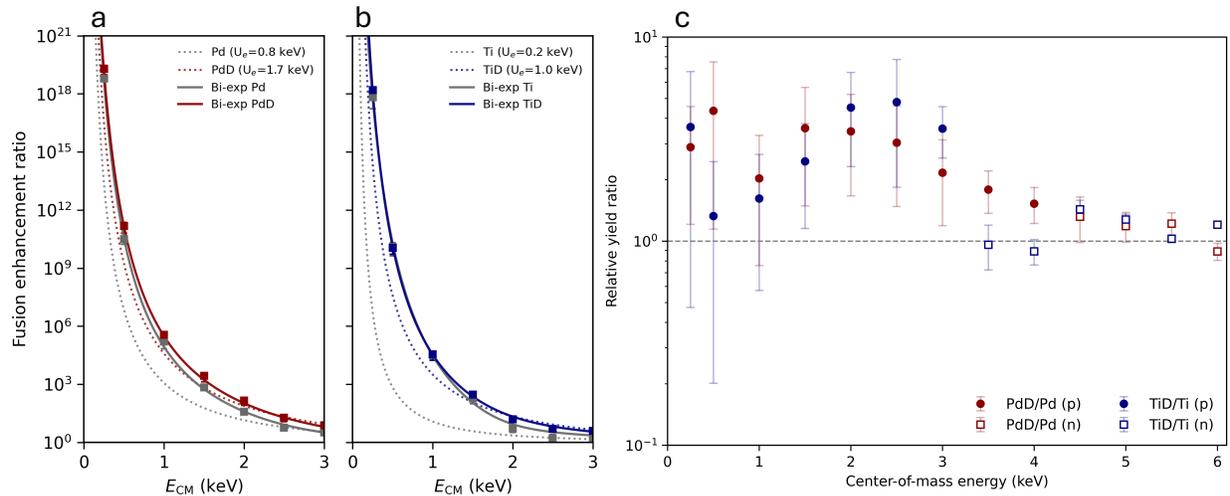

**Figure 3 | Fusion enhancement analysis and hydride–metal yield ratios. (a, b)** Enhancement factor for (a) Pd, PdD and (b) Ti, TiD as a function of center-of-mass energy. Dotted curves are the single-exponential model built from constant screening potentials. Across both materials, the single-exponential form fails to capture the measured curvature in this energy range, whereas a bi-exponential fit provides better agreement across both regions. **(c)** Hydride-to-metal relative-yield ratio for Pd (red) and Ti (blue) showing that electrochemical loading increases the yield for both materials across all energies probed in both proton (p) and neutron (n) reaction channels. Error bars denote 1σ counting uncertainties propagated through the yield normalization.

# Discussion

Both electrochemically driven hydrides (PdD and TiD) show enhanced electron screening compared to their non-electrochemically loaded counterparts (Pd and Ti), resulting in increased fusion yields. Electron screening is thought to arise from both free charge carriers in the material, as well as localized charge variations near the deuterons. Palladium remains metallic upon hydrogen uptake with a substantial density of states at the Fermi level, whereas Ti and its hydrides are less conductive with fewer itinerant carriers—consistent with a weaker screening response[26-28]. However, we observe that upon hydrogenation, both hydrides experience increased screening. One potential explanation for this effect is that electrochemical loading inserts deuterium into lattice interstitials, which locally alters the electromagnetic environment around those sites. When the deuterium atom takes its place within the interstitial site, its 1s orbital hybridizes with the valence bands of the metal, drawing in additional charge density in the vicinity of the deuteron[29]. As a result, a partially negative, polarizable D–metal bonding state is formed, which may locally enable increased screening. In addition, lattice defects (vacancies, dislocations) can trap deuterium, increasing the local density of reacting pairs and further modify the electronic environment[20,30,31]. Under elevated hydrogen chemical potential, Pd can stabilize superabundant-vacancy states that alter both the lattice and the electronic structure, even introducing D–D states in vacancy complexes[21,32]. Low energy ion implantation further creates near-surface defects and interstitials, compounding the defect density[33].

The observation of a plateau in D–D fusion below 2.5 keV is surprising. Ordinarily, deuterium fusion yields are expected to decrease monotonically with decreasing reaction energy. One possibility is that the screening effect varies with deuterium ion energy as ions probe different depths near the surface of the metal foil targets[34]. The defect density and deuterium loading can vary significantly with depth from the surface, leading to different electron screening environments and hence tunneling probabilities. Because the measured fusion yield includes contributions from different depth regions, $U_e$ is expected to be a function of depth and, in turn, ion energy. To capture this dependence, a simple bi-exponential model is found to be consistent with the experimental data over the measurement range. Alternatively, it has been proposed that a plateau in fusion rate could arise if there are additional low energy nuclear resonances[35,36]. Experiments at lower ion beam energy can further probe this regime.

While an electron screening model has been employed to describe fusion rates in metal targets, other effects could also be at play. Deuterium diffuses more rapidly in Pd than Ti, allowing it to move more freely and replace vacancies or depleted regions within the foil more readily than Ti. Under ion beam irradiation, the near-surface deuterium concentration reflects a balance between beam implantation, thermally and ion-stimulated losses to vacuum, and diffusive replenishment from the bulk during electrochemical loading. Because Pd exhibits much higher hydrogen mobility than Ti, Pd can more readily sustain the near-surface population required for low-energy reactions under otherwise similar loss fluxes, whereas Ti could develop a relatively depleted layer when diffusion and de-trapping cannot keep pace. These properties may also play a role in the different fusion yields for these materials. The results of several effects may be lumped into one parameter, the effective screening potential model $U_e$, even though they are physically different mechanisms.

# Conclusion

This work establishes a new experimental regime in which nuclear fusion at sub-keV energies can be reproducibly controlled by the materials environment. By coupling electrochemical deuterium loading with low-energy ion beam excitation, we uncover a plateau in D–D fusion yield below 2.5 keV that defies the expected exponential suppression. By comparing relative thick-target yields from Pd and Ti foils with and without electrochemical loading, we extract screening potentials of 1.7 ± 0.2 keV for PdD and 0.9 ± 0.2 keV for TiD—substantially higher than their non-electrochemically loaded counterparts, while suggesting that the local electronic and defect structures of metals actively shape tunneling probabilities.

The fusion enhancements observed here, reaching over $10^{18}$ times the bare-nucleus rate at an ion beam energy of 0.25 keV, demonstrate that nuclear reaction probabilities can be strongly modulated through materials design. Because both Pd and Ti show similar trends despite their distinct electronic and structural properties, the effect is likely general across metal hydrides and alloys, providing a foundation for systematic exploration.

These findings demonstrate a controllable platform for fundamental studies of nuclear processes in solids. They provide a basis for mapping the dependence of reaction rates on composition, defect density and hydrogen transport properties, opening a rational design space of materials that can modulate nuclear processes. Future experiments can help delineate microscopic mechanisms with experiments that can include tests with ion beams other than deuterium and other excitation sources including photons, THz phonons, and plasmons[37]. More broadly, this work opens a path toward the rational design of condensed-matter systems that can modulate nuclear reaction rates—linking the physics of fusion with the chemistry and microstructure of the host solid. While the fusion probabilities demonstrated here are many orders of magnitude below energy-production thresholds, the ability to tune nuclear reaction environments through materials design may ultimately inform aspects of long-term fusion-energy research, particularly in understanding how solid-state conditions alter reaction pathways.

# Acknowledgments:

This work was funded by the Advanced Research Projects Agency-Energy (ARPA-E), U.S. Department of Energy, under Award Number 22/CJ000/04/01. Work at Lawrence Berkeley National Laboratory was conducted under US Department of Energy contract DE-AC02-05CH11231. This material is based upon work supported by the National Science Foundation Graduate Research Fellowship under Grant No. 2036201 (MEK). We thank Takeshi Katayanagi for his support at Lawrence Berkeley National Laboratory where we conducted the beam experiments for this study. We thank Igor Jovanovic and Valentin Fondement for their insights and discussions of the CAEN data acquisition system. We thank Curtis Berlinguette, Yet-Ming Chiang, David K. Fork, Ross Koningstein, and Matthew D. Trevithick for many fruitful discussions, and Phil Schauer for the early joint work on the experimental concept and reactor design.

# Author Contributions:

MEK contributed to the experimental design and build of the setup, conducted all experiments and materials characterization, data analysis, and manuscript preparation. MC contributed to building the setup and data collection. AP contributed to building the setup and supported data interpretation. TS contributed to the conceptual and experimental design, contributed to setup construction, contributed to project management and manuscript writing, supervised the study, defined the hypotheses, and designed the project. JNM contributed to project management, experimental design and hypotheses, supervised the study, and contributed to drafting and revising the manuscript. All authors discussed the results and approved the final manuscript.

# Citations